\documentstyle[12pt]{article}
\title{Notes on covariant quantities in noninertial frames 
and invariance of radiation in classical and quantum field theory}
\author{Hrvoje Nikoli\'c  \\
Theoretical Physics Division, Rudjer Bo\v{s}kovi\'{c} Institute, \\
P.O.B. 1016, HR-10001 Zagreb, Croatia \\
{\normalsize hrvoje@faust.irb.hr} \\
\makebox[1in]{} \\
}
\date{\today}
\begin{document}
\maketitle
\begin{abstract}
A local observer can measure only  
the values of fields at the point of his own position.    
By exploring the coordinate transformation between two 
Fermi frames, it is shown that two observers, having the same instantaneous 
position and velocity, will observe the same values of covariant fields 
at their common instantaneous position, even if they have different 
instantaneous accelerations. In particular, 
this implies that in classical physics 
the notion of radiation is observer independent, contrary to the conclusion
of some existing papers. A ``freely" falling charge in  
curved spacetime does not move along a geodesic and therefore 
radiates. The essential feature of the Unruh effect is the fact that 
it is based on a noninstantaneous measurement, which may also be 
viewed as a source of effective noncovariance of measured quantities. 
The particle concept in Minkowski spacetime is clarified. It 
is argued that the particle concept in general spacetime 
does not depend on the observer and that there exists a preferred 
coordinate frame with respect to which the particle number 
should be defined.       
\end{abstract}
\vspace*{0.5cm} 
PACS number(s): 04.20.Cv, 03.50.-z, 41.60.-m, 04.62.+v

\section{Introduction}

As is well known, an accelerated charge in flat Minkowski spacetime 
radiates, at least as seen by an inertial observer. This naturally 
raises the following questions: Does an inertially moving charge 
radiate from the point of view of an accelerated observer? Does 
an accelerated charge radiate from the point of view of an observer 
who also accelerates with the same acceleration? Can one 
generalize the answers to these questions to curved spacetime 
and gravitational acceleration, only using the equivalence principle? 
The first aim of this paper is to give the answers to these and some related 
questions in the framework of classical physics.  

These questions have already been discussed in several papers 
[1-9]. In \cite{kovetz,pauri} the fields of an 
inertial charge in flat spacetime are transformed to Rindler 
coordinates, in which the Poynting vector ${\bf S}=
{\bf E}\times {\bf B}$ does not vanish everywhere. It is therefore 
concluded that an inertially moving charge radiates 
from the point of view of an accelerated observer.  
Similarly, by transforming the fields of a uniformly 
accelerated charge expressed in Minkowski coordinates   
to Rindler coaccelerating coordinates, it is found 
that the Poynting vector vanishes at all points to which 
Rindler coordinates can be applied. It is therefore concluded 
in \cite{bondi,rohr2,kovetz,boul,pauri} that the 
accelerated charge does not radiate from the point of view of an 
coaccelerating observer, although it does radiate from the 
point of view of an inertial observer. In \cite{bondi,boul,bell} 
it is also argued that a noninertial static charge 
in a static gravitational field does not radiate from the 
point of view of a static observer, because (since everything is 
static in the corresponding coordinates) the fields must be time independent.  

We criticize such a reasoning. 
For a given coordinate frame, the values of covariant 
fields are formally determined at 
all spacetime points covered by these coordinates.  
However, a local observer can measure only 
the values of fields at the point of his own position. It is completely 
unphysical to talk about the value of a field at some point 
from the point of view of the observer sitting at some other 
point. Rindler coordinates are a special case of Fermi coordinates.  
By exploring the coordinate transformation between two
Fermi frames, we show that two observers, having the same instantaneous
position and velocity, will observe the same values of covariant fields
at their common instantaneous position, even if they have different
instantaneous accelerations. When radiation is defined in an  
operationally meaningful way, this fact leads to the conclusion 
that radiation does not depend on the observer, but only on 
the motion of the radiating charge.  
    
We also study the concept of particle in quantum field theory in 
flat and curved spacetime. It is often argued 
(originally in \cite{unruh}) that a response of an 
accelerated ``particle detector" can be interpreted as a 
dependence of the particle number on coordinates. Although one 
could even accept that Minkowski coordinates 
have a privileged role in flat spacetime, it is usually argued that there is 
no such privileged coordinate frame in general curved spacetime. 
In this way 
the concept of particle becomes very fuzzy and, in particular, 
observer dependent. 

We argue, both formally and operationally, that the number of 
particles in a given quantum state does {\em not} depend 
on the observer. We give an invariant operational meaning to the 
particle number by discussing the response of classical detectors, 
such as a Wilson chamber or a Geiger-M\"{u}ller counter. 
We also argue that quantum physics and cosmological observations 
imply that a preferred coordinate frame {\em does} exist, which leads 
to a preferred representation of the algebra of quantum fields, but   
does not violate the covariance. This preferred representation 
leads to an unambiguous formal definition of the particle number.      

The paper is organized as follows: In Sec. 2 we 
explain the physical meaning of Fermi coordinates and study 
the coordinate 
transformation between two Fermi frames. In Sec. 3 we give a general 
discussion on measurable quantities in classical field theory. The 
concept of classical radiation is discussed in Sec. 4, where it is also 
shown that a ``freely" falling charge in        
curved spacetime does not move along a geodesic and consequently 
radiates. In Sec. 5 we argue that in quantum field theory the 
measurable quantities do not need to transform covariantly, which 
illuminates the origin of the Unruh effect. We also argue that the essential 
property of the measurements that can lead to the Unruh and similar 
effects is a large duration of the measuring procedure. The concept 
of particle in Minkowski spacetime is clarified in Sec. 6.  
Section 7 is devoted to the concept of particle in general 
spacetime, while Sec. 8 is devoted to the concluding remarks. 

We use the signature $(+---)$ and the units $\hbar =c=k_{B}=1$, where 
$k_{B}$ is the Boltzmann constant.

\section{Coordinate transformation between two Fermi frames}

Before starting with calculations, we clarify the physical 
meaning of Fermi coordinates, reviewing some results  
already presented in \cite{nikolic1}. Fermi coordinates are 
proper coordinates 
of a local observer determined by its trajectory in spacetime. 
It is convenient to define them such that the observer is 
positioned at the space origin. Therefore, the metric expressed 
in Fermi coordinates has the property 
$g_{\mu\nu}(t,0,0,0)=\eta_{\mu\nu}$ \cite{mtw}.
Even if there is no relative motion between two observers, they 
belong to different Fermi frames if their trajectories do not 
coincide. However, for practical purposes, they can be considered 
as belonging to the same frame if there is no relative motion between 
them and the other observer is close 
enough to the first one, in the 
sense that the metric expressed in Fermi coordinates 
of the first observer does not depart  
significantly from $\eta_{\mu\nu}$ at the position of the second 
one. In particular, this implies that two inertial observers in 
flat spacetime who have the same velocity can always be considered 
to be in the same frame, no matter how distant they are from each other. 
In other words,  
Fermi coordinates have a clear physical interpretation only in a
small neighborhood of the physical observer to whom the Fermi
coordinates refer. It is an exclusive
property of Minkowski coordinates, among other inertial and
noninertial Fermi coordinates, to have a clear global
physical interpretation. Since the intuition of physicists is 
often based on the familiarity with Minkowski coordinates, 
this is the source of many misinterpretations.  
Some of them have been discussed in \cite{nikolic1}, 
while some others are discussed in this paper.      

We first consider the coordinate transformation between different 
Fermi frames in flat spacetime. 
Let $S$ be an inertial frame
and let $S'$ be the frame of the observer whose 3-velocity is 
$u^i(t')\equiv \mbox{\bf{u}}(t')$, as seen by an observer in $S$.
In general, $S'$ can also rotate, which can be described by 
the rotation matrix $A_{ji}(t')=-A_{j}^{\; i}(t')$.  
The coordinate transformation between these two frames can 
be written as \cite{nikolic1}
\begin{equation}\label{el3}
x^{\mu}=\int_{0}^{t'}f^{\mu}_{\; 0} (t',0;\mbox{\bf{u}}(t')) dt' +
\int_{C}
f^{\mu}_{\; i} (t',\mbox{\bf{x}}';\mbox{\bf{u}}(t')) dx'^{i} \; ,
\end{equation}
where
\begin{equation}\label{partial}
f^{\mu}_{\; \nu}=\left( \frac{\partial f^{\mu}}{\partial x'^{\nu}} 
 \right)_{\mbox{\bf{u}}=\mbox{\rm{const}}} \; , 
\end{equation}
and   
\begin{equation}\label{el1}
x^{\mu}=f^{\mu}(t',\mbox{\bf{x}}';\mbox{\bf{u}})
\end{equation}
denotes the ordinary Lorentz transformation, i.e., the transformation
between two inertial frames specified by the constant relative velocity
$\mbox{\bf{u}}$. In (\ref{el3}), $C$ is an arbitrary curve with constant
$t'$, starting from $0$
and ending at $-A_{j}^{\; i}(t')x'^{j}$. It is assumed that 
the space origins of $S$ and $S'$ coincide at $t=t'=0$. The coordinates 
$x'^{\mu}$ are the Fermi coordinates of the observer positioned at $x'^{i}=0$. 
By combining two transformtions of the form (\ref{el3}) one can 
find the transformation between two noninertial Fermi frames. The 
quantity $f^{\mu}_{\; \nu}=\partial x^{\mu}/ \partial x'^{\nu}$, 
which is relevant to  
the transformation of tensors, is, in general, a complicated function 
of $x'$. However, if one restricts the analysis to the point $x'^{i}=0$, 
one obtains much simpler relations. 

Let as calculate $f^{\mu}_{\; \nu}$ at $x'^{i}=0$ for a 
{\em fixed} instant $t'$. Without losing on generality, we choose the rotation 
matrix such that $A_{ij}(t')=\delta_{ij}$, which means that the 
corresponding space axes 
of $S$ and $S'$ are parallel at this particular instant.   
From (\ref{el3}) and the well-known Lorentz transformations in (\ref{el1}) 
one easily finds   
\begin{eqnarray}\label{fmini}
& f^{0}_{\; 0}=\gamma \; , \;\;\;\;\; f^{0}_{\; j}=-\gamma u_j \; , \;\;\;\;\;
  f^{i}_{\; 0}=\gamma u^i \; , & \nonumber \\
& f^{i}_{\; j}=\delta^{i}_{\; j}+
 \displaystyle\frac{1-\gamma}{\mbox{\bf{u}}^2} u^i u_j \; , &
\end{eqnarray}
where $u^j =-u_j$, $\mbox{\bf{u}}^2 =u^i u^i$,
$\gamma=1/\sqrt{1-\mbox{\bf{u}}^2}$ are evaluated at $t'$. We see 
that $f^{\mu}_{\; \nu}$ at the position of the observer in $S'$ 
depends only on the instantaneous velocity and that this dependence 
is the same as for two inertially moving observers. Actually, 
in a more general case there is 
also a dependence on the instantaneous orientation of the 
space axes, but this is physically irrelevant.    

Let us now generalize these results to curved spacetime. 
Let $S$ and $S'$ be the Fermi frames 
of two arbitrarily moving observers. Assume that they have 
the same position at $t=t'=0$. This implies that the 
coordinate transformation between the two coordinate frames 
takes the form
\begin{equation}\label{transf}
x^{\mu}=f^{\mu}_{\; \nu} \, 
x'^{\nu} + f^{\mu}_{\; \nu\alpha}x'^{\nu}x'^{\alpha} 
+ \ldots \; .
\end{equation}
We are interested only in the quantity
\begin{equation}\label{fmunu}
f^{\mu}_{\; \nu}=\left( \frac{\partial x^{\mu}}{\partial x'^{\nu}}
 \right)_{x=x'=0} \; .
\end{equation} 
The metric tensor at $x=x'=0$ transforms as 
\begin{equation}\label{metric}
g'_{\mu\nu}=f^{\alpha}_{\; \mu}f^{\beta}_{\; \nu}\, g_{\alpha\beta} \; .
\end{equation}
Since we study the Fermi coordinates, the metric has a form
$g_{\alpha\beta}(x)=\eta_{\alpha\beta} +{\cal O}(x^i)$, 
$g'_{\mu\nu}(x')=\eta_{\mu\nu} +{\cal O}(x'^i)$. Putting this in 
(\ref{metric}), we obtain the equation
\begin{equation}\label{metric2}
\eta_{\mu\nu}=f^{\alpha}_{\; \mu}f^{\beta}_{\; \nu} \, 
\eta_{\alpha\beta} \; .
\end{equation}
Therefore, the problem of finding the coefficients $f^{\mu}_{\; \nu}$ 
reduces to the problem of finding the coefficients of a 
linear transformation $x^{\mu}=f^{\mu}_{\; \nu} \, x'^{\nu}$ that 
preserves the Minkowski metric $\eta_{\mu\nu}$. This is nothing else 
but how the ordinary Lorentz transformations are usually found. The solution 
of this problem is well known and is given by (\ref{fmini}), 
where $u^i=dx^i/dt$ is the velocity of the observer in $S'$ as 
seen by the observer in $S$, at the instant when the two observers 
have the same position. Of course, one can generalize (\ref{fmini}) 
for the case when one of the observers rotates relative to the other, 
but this is again physically irrelevant, because one can always 
choose the space axes of the two frames such that the 
corresponding axes are parallel at the instant of interest. 

The physical consequence of the result obtained in this section 
is the following: Let $\Phi_{\alpha_1 \ldots \alpha_n}(x)$ be 
an arbitrary local tensor quantity. Let the two observers 
measure this quantity at their common instantaneos position. 
The results of measurements will be related as 
\begin{equation}\label{Phi}
\Phi'_{\mu_1 \ldots \mu_n}=f^{\alpha_1}_{\; \mu_1}
 \cdots f^{\alpha_n}_{\; \mu_n} \, \Phi_{\alpha_1 \ldots \alpha_n} \; .
\end{equation}
The two measurements will be different if there is an instantaneous 
relative velocity between the two observers. However, the 
instantaneous relative acceleration, as well as higher-order 
derivatives, are irrelevant to this transformation law.        

Equation (\ref{Phi}) is the local transformation law of 
covariant tensors. One can easily find the generalization 
for contravariant or mixed tensors, by performing contractions 
with $\eta^{\mu\nu}$. 

\section{What is measurable in classical field theory?} 
 
In this section we explain in more detail the local 
nature of classical measurements. 

A local observer can measure only
the values of fields at the point of his own position. 
It is completely
unphysical to talk about the value of a field at some point
from the point of view of the observer sitting at some other
point. For example, when somebody ``sees" a distant object, 
he actually measures the properties of reflected light at his own 
position. Since all the interactions are local, a measuring 
apparatus can only response to the values of fields at the position 
of the apparatus. 

In practice, a measuring apparatus is usually a large object. 
In this case, each part of the apparatus should be regarded 
as a separate local measuring device, each having its own 
Fermi coordinates. The response of the apparatus as a whole 
is some kind of sum or average of the responses 
of all its parts, where the interaction between various 
parts of the apparatus may also play a significant role.
In this sense, a measurement made using a 
large apparatus can be regarded as an unideal measurement. 
Some similar remarks have already been given  
in \cite{parrott}, but the consequences of such a reasoning 
have not been explored completely. 

To calculate how a large measuring apparatus will respond to a 
given field at a certain instant (or, better to say, at 
a certain spacelike hypersurface), one needs to specify the 
velocity of each part of the apparatus at this instant. 
If the measured field transforms as a tensor, the 
instantaneous accelerations are irrelevant. 

A measurement can also be unideal if it lasts a finite time. 
Obviously, if two observers have equal velocities initially, but 
different velocities later, a cummulative effect,  
which depends on all instantaneous velocities, will 
be different for the two observers. However, it is important 
to realize that the acceleration itself is not essential 
for understanding such effects. (An example of such an 
effect is the twin paradox, where, contrary to the common 
belief, the acceleration is not essential. For example, 
in curved spacetime it may be possible to connect two points 
by two different timelike geodesics that have 
different proper lengths. If these two 
geodesics correspond to the trajectories of two observers, one 
will obtain the twin ``paradox" without acceleration.) 

In practice, no measurement is ideal. However, we can 
introduce the concept of almost ideal measurements, i.e., 
the measurements that  
last short enough such that the velocity and the measured 
field do not change significantly and are performed 
using a small apparatus such that the metric of the corresponding 
Fermi coordinates and the measured field do not vary significantly 
through the apparatus.  
    
\section{What is radiation in classical electrodynamics?}

In this section we attempt to give an operational and 
therefore physically meaningful definition of the 
concept of radiation in classical electrodynamics, based 
on ideal or almost ideal measurements. 

In \cite{rohr1} the measure of radiation is defined 
in a Lorentz invariant manner, as the 
flux of the Poynting vector through a closed two-dimensional 
surface that looks as a sphere for an inertial 
observer instantaneously comoving with the charge at 
the retarded time. It is also shown that the radius 
of this sphere does not need to be large. We call this 
the standard definition of radiation. 
 
However, such a definition of radiation is 
global. We need criteria that can 
answer the question whether there is a radiation at 
a certain spacetime point. We also require that 
it should be possible to apply the criteria 
to inertial 
as well as to accelerated observers, in flat as well as in 
curved spacetime. 

There is an attempt \cite{mould} to show that an inertial
charge in flat spacetime radiates from the point of view    
of an accelerated observer, by studying the response of a 
specific model of a classical detector. It is shown that this detector
absorbs energy when accelerates uniformly, but does not absorb 
energy when moves inertially.
However, the behavior 
of this specific unideal ``detector" proves nothing, because
one can easily model a detector that will absorb energy 
even when moves inertially. If one wants to see whether the
concept of radiation depends on the motion of the observer, 
the answer should not depend on details of the measuring 
apparatus.   

We note that radiation is not a kinematical effect 
resulting from the coordinate transformation between 
the frames of the radiating charge and the observer, but a 
dynamical effect, in the sense that even for the observer 
comoving with the charge, the fields depend on acceleration. 
This is not explicitly seen in the conventional approach 
in which the Maxwell equations are solved in 
Minkowski coordinates. To see  
this explicitly, we write the covariant Maxwell equation 
\begin{equation}\label{max1}
D_{\mu}F^{\mu\nu}=j^{\nu} 
\end{equation}
in a more explicit form 
\begin{equation}\label{max2}
\partial_{\mu}F^{\mu\nu}+\Gamma^{\mu}_{\mu\lambda}F^{\lambda\nu}
 +\Gamma^{\nu}_{\mu\lambda}F^{\mu\lambda}=j^{\nu} \; .
\end{equation}  
We assume that the current $j^{\nu}$ corresponds to a pointlike 
charge. Let us study how the field $F^{\mu\nu}$ looks like to 
an observer comoving with the charge in his small neighborhood. 
Since he uses the corresponding Fermi coordinates, 
the connections $\Gamma^{\alpha}_{\beta\gamma}$
vanish in his
small neighborhood if and only if his trajectory is a
geodesic \cite{mtw}. Therefore, 
if the charge does not accelerate, in the small neighborhood 
of the charge the solution of (\ref{max2}) looks just like 
the well-known Coulomb solution ${\bf E}\!\propto\! r^{-2}$, 
${\bf B}\! =\! 0$. 
On the other hand, if the charge accelerates, then, even in the 
small neighborhood, Eqs. (\ref{max2}) no longer look  
like the Maxwell equations in Minkowski spacetime. 
This gives rise to a more complicated solution, which includes the 
terms proportional to $r^{-1}$. A similar approach to radiation has 
been studied in \cite{padmg}. 
We can apply the 
standard definition of radiation to the small 
neighborhood of the charge and conclude in this way that the accelerated 
charge always radiates and the inertial charge never does, even in curved 
spacetime. In this way, the radiation is an intrinsic property 
of the fields in the small neighborhood of the radiating charge, 
seen by a comoving observer. If this intrinsic 
definition of radiation is 
accepted, then, obviously, radiation does not depend on the 
observer. 

Once the fields leave its source, their propagation is determined 
only by the spacetime geometry. Therefore, one should be able 
to answer the question whether the charge radiates by measuring 
the fields at arbitrary distances from the charge. However, the conclusion 
should not depend on the observer. 

One could try to 
propose that at a certain spacetime point there is radiation if 
the Poynting vector does not vanish at this point. However, even two  
inertially moving observers in flat spacetime may not agree on whether 
the Poynting vector vanishes at a certain point. 

One could propose that ``to radiate" means ``to emit energy". 
However, in non-Minkowski spacetime, the global concept 
of energy is not well defined. One could use our local 
philosophy to note that in the vicinity of any observer the 
metric is Minkowskian, so the energy that he can measure is 
well defined. However, there is a trouble again, because,  
even in globally Minkowskian spacetime, the 
energy-momentum tensor is defined up to a total derivative, 
which does not affect the total energy, but does 
affect the energy contained in a small volume. Nevertheless, 
one can choose some specific definition of the 
energy-momentum tensor, which, in our opinion, leads to 
the best possible definition of energy, namely, the 
energy contained in a small volume measured by a  
local observer. All other definitions of energy, such as  
the global energy based on a timelike Killing 
vector (if it exists), are unoperational and therefore 
unphysical.            

The only measurable entity related to the electromagnetic field 
is its effect on charges, described by the covariant equation 
\begin{equation}\label{force} 
m \ddot{x}^{\mu}=q \, F^{\mu\nu}_{{\rm ext}}\, \dot{x}_{\nu} +
\Gamma^{\mu} \; ,
\end{equation}
where 
\begin{equation}\label{selfforce}
\Gamma^{\mu}=\frac{2}{3}\, q^2 \left[ \ddot{v}^{\mu} -v^{\mu} 
\dot{v}_{\alpha}\dot{v}^{\alpha} \right] +
\Gamma^{\mu}_{{\rm curv}} \; 
\end{equation}
and the quantities
\begin{eqnarray}\label{derivs}
& v^{\mu}=\dot{x}^{\mu}=d x^{\mu}/d\tau \; , & \nonumber \\
& \dot{v}^{\mu}=\ddot{x}^{\mu}=d \dot{x}^{\mu}/d\tau +
  \Gamma^{\mu}_{\beta\gamma}\dot{x}^{\beta}\dot{x}^{\gamma} \; , & 
  \nonumber \\
& \ddot{v}^{\mu}= d \ddot{x}^{\mu}/d\tau +
  \Gamma^{\mu}_{\beta\gamma}\ddot{x}^{\beta}\dot{x}^{\gamma} & 
\end{eqnarray}
transform as vectors \cite{witt1}. In (\ref{force}) 
the first term represents the familiar Lorentz force caused by an  
external field $F^{\mu\nu}_{{\rm ext}}$. The $\Gamma^{\mu}$ term  
represents the self-force, where the first term in (\ref{selfforce}) 
is the well-known Abraham-Lorentz force. The 
$\Gamma^{\mu}_{{\rm curv}}$ term depends on the curvature and vanishes  
when the curvature is zero, but for the nonvanishing curvature 
it is nonvanishing even when the acceleration $\dot{v}^{\mu}$ and 
the higher derivatives are zero. The explicit form of 
$\Gamma^{\mu}_{{\rm curv}}$ is given in \cite{hobbs,qwald}. 

There is a simple intuitive picture explaining why a self-force 
appears when the charge accelerates or when spacetime around the charge  
is curved. The fields produced by the charge always act on it. However, 
when the charge moves inertially through flat spacetime, then 
the metric related to the corresponding Fermi coordinates is 
isotropic, 
and so are the fields. This implies that the self-forces 
in different directions cancel exactly, so the resultant force 
is zero. When spacetime is curved (such that it is not isotropic) 
or the charge accelerates, then 
the metric related to the Fermi coordinates is no longer 
isotropic. Consequently, the fields are also not isotropic, 
which implies 
that the resultant force need not to be zero. 
 
The presence of the $\Gamma^{\mu}_{{\rm curv}}$ term in (\ref{selfforce}) 
implies that even in the absence of external electromagnetic fields, 
a charge in curved spacetime does not move geodesically. This is 
not inconsistent with the equivalence principle, because it 
states that when {\em only} 
gravitational fields act on a pointlike particle, 
then the motion of the particle does not depend on internal properties 
of the particle (such as its charge). In our case, there are also 
electromagnetic forces produced by the charge. Alternatively, one can 
regard the charge and its fields as one object, which is no longer 
pointlike, so the assumptions of the equivalence principle are  
violated again.   
    
Since a ``freely" falling charge 
in curved spacetime does not move geodesically, 
it follows, according 
to our intrinsic definition of radiation and Eqs. (\ref{max2}), 
that it radiates. Of course, in curved spacetime one does not 
expect, in general, that radiating fields will fall off as $r^{-1}$ 
at large distances (it is also not clear what $r$ is in curved 
spacetime). However, it is reasonable to expect that at large distances 
the fields produced by accelerated (i.e., nongeodesically moving) 
charges will be much stronger than those of geodesically moving charges. 

Now we turn back to the attempt to give an operational definition 
of radiation at large distances. In our opinion, 
the only reason why radiating fields deserve special attention 
in physics, is the fact that they fall off 
much slower than other fields, so their effect is 
much stronger at large distances. 
Actually, the distinction between ``radiating" and 
``nonradiating" fields is quite artificial; there is only one field, 
which can be written as a sum of components that fall off differently 
at large distances. If one knows the distance of the charge 
that produced the electromagnetic field $F^{\mu\nu}_{{\rm ext}}$ 
and measures the intensity of its effects described by (\ref{force}), 
then one can determine whether this effect is ``large" or ``small", i.e., 
whether the charge radiates or not. If the field is proportional 
to $r^{-1}$ as seen by one observer, it is also so as seen by any other 
observer at the same position. Since $F^{\mu\nu}_{{\rm ext}}$ in 
(\ref{force}) transforms as a tensor, the effect may depend on the  
velocity of the observer, but cannot depend on its acceleration. 
In this sense, we can say that radiation does not depend on 
the observer. 

To conclude this section, we note that the concepts of energy and 
radiation in classical field theory are only auxiliary concepts. 
These concepts may be useful and well defined in Minkowski 
spacetime, but not in general.  
All that one really needs is contained in unambiguous equations of 
motions, such as (\ref{max1}) and (\ref{force}). However, for 
those who insist on giving a definition of these quantities in 
general spacetime, in this section we have attempted to give 
the best possible operational definitions. One may feel that 
these definitions are far from being precise, but this is because 
we have attempted to define concepts which are not really 
meaningful at the fundamental level. Nevertheless, it is our 
conclusion that the best possible definition of radiation leads to 
the conclusion that 
the charge ``radiates" if and only if it does not 
move geodesically. It does not depend on the observer.  
  
\section{Noncovariant quantities, time nonlocal measurements, 
and quantum field theory}    
  
If a measurable quantity does not transform covariantly, i.e., as a 
tensor, then the transformation law (\ref{Phi}) cannot be applied. 
This fact can be particularly illuminating for understanding  
some effects of quantum field theory. For example, if one does not 
use the normal ordering of operators, then the 
expectation value of the energy-momentum tensor  
can always be written as  
\begin{equation}\label{tmini}  
\langle\psi | T^{\mu\nu} | \psi \rangle = t^{\mu\nu}_{{\rm inf}}-
t^{\mu\nu}_{{\rm meas}} \; ,
\end{equation}
where $t^{\mu\nu}_{{\rm inf}}$ is an infinite unmeasurable part, 
whereas $t^{\mu\nu}_{{\rm meas}}$ is the 
measurable part. (The minus sign in (\ref{tmini}) 
is a matter of convenience; one can also obtain a positive sign by 
redefining the infinite part.) 
The left-hand side formally transforms 
covariantly, but it is infinite and therefore cannot be measurable. 
The measurable 
part depends on details of the measuring procedure. 
Accordingly, there is no 
{\it a priori} reason why the two terms on the right-hand side 
should transform separately as tensors. 

The Unruh effect can be viewed as an example of such an 
effect. If $|\psi\rangle =|0\rangle$, then  
$t^{\mu\nu}_{{\rm meas}} =0$ for an inertial observer. On the other hand,
an observer accelerated uniformly with the acceleration $a$ 
measures a thermal energy-momentum tensor corresponding to the 
temperature $T=a/2\pi$, which does not vanish. Obviously, the 
measurable part of $\langle 0| T^{\mu\nu} |0 \rangle$ 
does not transform covariantly. To see more closely how this 
happens, consider a real massless scalar field in $1+1$ dimensions 
quantized using the standard Minkowski quantization. We study the 
correlation function
\begin{equation}\label{corel}
\Gamma(x_i,x_f)=\frac{1}{4}\, 
\langle 0|\partial_0\phi(x_i)\partial_0\phi(x_f) +
\partial_1\phi(x_i)\partial_1\phi(x_f) +(x_i \leftrightarrow x_f) 
|0\rangle \; , 
\end{equation}
where $x_i=x(\tau_i)$, $x_f=x(\tau_f)$, and $x(\tau)$ is the trajectory 
of the observer. The partial derivatives are calculated with respect 
to the Fermi coordinates of the observer. This correlation function has 
the property
\begin{equation}\label{simt00}
\langle 0|S\, T^{00}(x)|0\rangle =\Gamma(x,x) \; ,
\end{equation}
where $S$ denotes the symmetric ordering. Using the results of 
\cite{meyer}, we can immediately write (\ref{corel}) for a 
uniformly accelerated observer. It appears that $\Gamma$ depends only 
on $\Delta\tau =\tau_f -\tau_i$ and can be written as 
\begin{equation}\label{corel2}
\Gamma(\Delta\tau)=\frac{1}{(\Delta\tau)^2}-\left[ \frac{1}{(\Delta\tau)^2}
- \frac{a^2}{4\, {\rm sh}^2 (a\,\Delta\tau /2)} \right] \; .
\end{equation}     
The first term, infinite in the limit $\Delta\tau\rightarrow 0$, 
corresponds to that which would be obtained for an inertial 
observer. The term in the square brackets vanishes for
$\Delta\tau\rightarrow 0$ and has a thermal Fourier transform \cite{meyer}. 
This suggests that the first term should be interpreted as the 
infinite unmeasurable term, while the term in the square brackets is 
the finite measurable one. This also demonstrates that this noncovariant 
effect disappears when the measurement is instantaneous, i.e.,
$\Delta\tau\rightarrow 0$.  

Note that the breaking of scale invariance through the 
renormalization in quantum field theory, well known by 
particle physicists, can also be viewed in a similar way. 
An infinite, but scale-invariant correlation function obtained 
before the renormalization can be written as a sum of two 
scale-noninvariant terms, one of them being the finite  
measurable term.  

To understand what is measurable and how the result depends on 
the duration of the measurement, one needs to study a model 
of the detector. In a simple pointlike ``particle detector" 
moving arbitrarily in flat spacetime and initially being 
in its ground state $E_0$, the amplitude 
for the transition to the excited state $E$ is \cite{bd} (see also 
\cite{padm1} for a justification of such a simple model) 
\begin{equation}\label{ampl}
A( |0\rangle\rightarrow |k\rangle , \; E_0 \rightarrow E)\propto 
\int_{\tau_i}^{\tau_f} d\tau \, e^{i(E-E_0 )\tau} \langle k|\phi(x(\tau))
|0\rangle \; , 
\end{equation}
where it is assumed that the field $\phi(x)$ is initially in its 
vacuum state $|0\rangle$ and $\tau_f -\tau_i$ is the duration of the
measuring procedure. The final state of the field is a 
one-particle state $|k\rangle$.  
In this model, the detector can spontaneuosly jump into its excited 
state, owing to vacuum fluctuations.   
Note that this amplitude is nonvanishing even for an inertially moving 
detector. Only if one takes $\tau_f -\tau_i =\infty$ 
(as is usual in quantum mechanics),  
one obtains a vanishing amplitude for an 
inertial detector and a thermal amplitude for a uniformly 
accelerating one. The ultimate reason
why the lowest energy state $|0\rangle$ appears as ``nothing"
to an inertial observer, although the energy of this 
state is actually infinite, is the fact that typical measurements                   
in quantum mechanics usually last very long compared with the
inverse energy differences measured. 
This is why one can say that the energy of vacuum is zero. 
One need not use the normal ordering in order to 
calculate the physical effects of various states. For example, 
when one uses the symmetric ordering of the Hamiltonian in 
perturbative calculations with infinite-time 
intervals and assumes that the field 
is in its vacuum state, one obtains two infinite terms that 
contribute to a spontaneous excitation of an atom in its 
ground state. However, these two terms cancel exactly for an inertial 
atom and lead to finite thermal excitations for a uniformly 
accelerating one \cite{muller}.  

The Unruh effect, viewed as an effect of vacuum fluctuations on 
a noninertially moving atom, is a quantum effect. However, it is 
important to note that the thermal nature of these excitations for 
a uniform acceleration 
can be understood even with classical physics \cite{padm2}. 
Moreover, even the effect of stochastic quantum fluctuations 
on correlation functions can be simulated by viewing vacuum as 
a sea of classical random fluctuations with a Lorentz invariant 
spectrum \cite{boyer}. 

It is also important to note that the actual response of a uniformly 
accelerating detector depends on details of interactions in the 
detector. In some cases, there is no response at all \cite{mur,marz}.
      
\section{The concept of particle in Minkowski spacetime}

The Unruh effect is often interpreted as absorption of a ``particle" 
which does not exist from the point of view of an inertial observer. 
Such an interpretation makes the concept of 
particle rather fuzzy. Before discussing the concept of particle  
in general spacetime in the next section, we find necessary 
first to clarify the concept of particle in an inertial, 
Minkowski frame. 

An $n$-particle state is defined as a normalized state of the form 
\begin{equation}\label{n-part}
|n\rangle =\int d^3 k_1 \cdots d^3 k_n \, 
f(k_1, \ldots ,k_n)\, a^{\dagger}(k_1) \cdots 
a^{\dagger}(k_n)|0\rangle \; .
\end{equation}
The crucial question we attempt to answer in this 
section is why such formally 
defined states correspond to that which is observed in 
typical experiments as $n$ separated entities, i.e., ``particles". 

We introduce the concept of a classical particle detector, such as 
a Wilson chamber or a Geiger-M\"{u}ller counter. We call such 
detectors classical, because, in order to understand how and why such
detectors respond, quantum field theory is not essential. 
Classical detectors respond to states that correspond to the classical 
concept of particle, i.e., to states which are well localized in 
space (small lumps of energy). For example, if two localized particles 
are very near each other, which can be achieved by a suitable choice 
of the function $f(k_1,k_2)$, 
a classical detector will see this state as one 
particle. However, it seems that such things do not occur in practice. Why?
 
Before answering this question, first consider how a one-particle 
state is detected. 
Assume that a particle detector is localized in space. The function 
$f(k_1)$ can correspond to a plane wave or to a state with two lumps, 
but such states cannot be observed by the localized detector. It will 
not be observed until the 
function $f(k_1)$ collapses to a one-lump state. (We do not attempt to 
answer the question whether the wave function collapses  
spontaneously or the detector causes it. In some interpretations of 
quantum mechanics there is a clear answer to this question, but we 
do not want to prefer any particular interpretation.) 

Now consider a two-particle state. Assume that, {\it a priori}, all 
functions $f(k_1,k_2)$ are equally probable. However, 
the number of functions corresponding  to a two-lump state is 
larger than that for a one-lump state, 
so it is very improbable that two particles will form 
one lump. (Of course, the numbers of two-lump states and one-lump states 
are both infinite, but their ratio is not equal to 1.) A four-lump state, 
for instance, is even more probable 
than a two-lump state, but if one particle is distributed 
in more than one lump, it will not be detected. 

In this way we have explained why an $n$-particle 
state behaves as an $n$-lump state in 
experiments. Of course, if there are strong attractive forces among 
fields, there may exist a natural tendency that $n$-particle states form 
a one-lump state (hadrons, $\alpha$-particles), in which case it is more 
convenient to treat such states as one-particle states. Actually, it is 
often completely incorrect to treat such states as $n$-particle states, 
because nonperturbative effects of interactions may completely 
change the spectrum of states of a free theory, such as is the case 
for QCD.     

One may argue that classical detectors are not the best operational   
way to define a particle. One should rather study quantum detectors.  
For example, an atom absorbs precisely one photon, not a half of it, nor 
two of them. Even the response of a ``classical" detector should be 
ultimately described in terms of virtual absorptions and 
emissions (this is the reason why a ``classical" localized 
detector cannot respond to a two-lump one-particle state). 
However, it is important to emphasize that there is not  
any deep, fundamental principle which forbids absorption or emission 
of a half or two particles. It is merely a consequence of a particular 
form of  
dynamics. For example, an electron in the atom absorbs {\em one} photon because 
the interaction Lagrangian ${\cal L}_I =e\bar{\psi}\gamma_{\mu}\psi A^{\mu}$ 
is {\em linear} in $A^{\mu}$. Of course, higher-order corrections allow 
absorption of two photons (sum of their energies must be equal to the 
difference of energies of the atom levels), but such processes are 
suppressed dynamically (small coupling constant) and kinematically (small 
probability of a one-lump two-photon state). But is it, in principle, 
possible to absorb a half of a particle? One can exclude such a 
possibility if one proposes that the Lagrangian ${\cal L}(\phi,
\partial_{\mu}\phi, \ldots )$ 
od fundamental fields 
must be an analytic function around zero. For example, 
assuming that there are no derivative couplings, the analyticity implies 
that a local interaction Lagrangian has a form 
\begin{equation}\label{lagr}
{\cal L}_I(\phi(x),\cdots ) =\sum_{n\geq 0}C_n(x)(\phi(x))^n \; , 
\end{equation}
where $n$ are integers and $C_n(x)$ depend on some other fields. 
The interaction (\ref{lagr})   
implies that the number of absorbed or emitted particles must be 
an integer. If $\phi(x)$ is an effective, composite field, like 
$\phi(x)=\chi(x)\chi(x)$, then one can have a term 
proportional to $\sqrt{\phi}=\chi$, in which case a ``half" of the 
$\phi$-particle, i.e., one $\chi$-particle, can be absorbed or emitted. 
If one allows terms like $\sqrt{\phi}$ for fundamental fields, 
then even a ``half" of a fundamental particle can be absorbed or emitted.  
(In this case, the concept of particle concept based on perturbative 
calculations 
is no longer a good concept. In particular, it is not clear, 
even algebraically, what $\sqrt{a^{\dagger}}|0\rangle$ is.)            
        
\section{The invariant concept of particle in general spacetime} 

In the preceding section we have seen that even in Minkowski spacetime 
the concept of particle in quantum field theory is not completely clear. 
This is, of course, hardly surprising, because already in classical 
physics the concepts of a field and of a particle are quite different. 
It is usually argued that in non-Minkowski spacetime the concept 
of particle is even more fuzzy (see, for example, \cite{bd} and 
references therein) and, in particular, that the number of particles 
in a given state depends on the motion of the observer. In this section 
we argue that the concept of particle in general spacetime is {\em not} 
more problematic than in Minkowski spacetime.      

One could be tempted to apply our local philosophy of Secs. 3 and 4, 
well suited for classical fields,  
to quantum fields as well. However, this would not be appropriate, because  
quantum physics possesses certain nonlocal properties, related to the 
uncertainty 
relations and EPR-like effects. Quantization requires a global 
approach, which is the reason that it is difficult to join quantum 
mechanics and general covariance.    

It is often claimed (originally in \cite{unruh2}) that the Unruh effect, 
discussed in Sec. 5, can be alternatively described by the so-called 
Rindler quantization, which consists in decomposing  free  
fields into modes proportional to $e^{\pm i\omega\tau}$, where $\tau$ is 
the time measured by a uniformly accelerated observer (this is the same time 
as $t'$ in (\ref{el3}) for a uniform acceleration). The corresponding 
number of ``Rindler particles" $N_R =\sum a^{\dagger}_R a_R$ in 
Minkowski vacuum $|0\rangle$ is not zero, but has a thermal 
distribution. Therefore, according to such an interpretation, the 
inertial and the accelerated observers do not agree on the number of 
particles contained in the state $|0\rangle$, and this is the reason 
that an accelerated atom can jump to an excited state. 

However, contrary to the common belief, the two descriptions of the  
Unruh effect are {\em not} equivalent. In particular, the 
Rindler-quantization approach predicts that the absorption 
of a Rindler particle by the accelerated atom will be seen 
by an inertial observer as an emission of a Minkowski particle 
\cite{unruh2} only if the atom has actually jumped to the 
excited state \cite{muller2}. 
On the other hand, by putting $E=E_0$ in (\ref{ampl}), 
we see that the Minkowski-quantization approach predicts an emission  
of a Minkowski particle even if the transition to an excited state 
has not actually occurred. The fact that the two inequivalent approaches 
both lead to a thermal spectrum with the temperature $T=a/2\pi$ is 
hardly surprising, because, as we have already noted, classical physics 
also leads to a thermal spectrum with the same temperature 
\cite{padm2,boyer}. And even this partial agreement of the two 
approaches does not generalize  
when the uniform acceleration is replaced by  
a more complicated motion \cite{padm3}. Since the two approaches 
are not equivalent, and since it is not reasonable to suspect 
that the predictions based on the Minkowski quantization are incorrect, 
one should reject the Rindler quantization as a correct description 
of nature.    

Let us also discuss it from a more formal point of view. 
In particular, we are interested in the question whether the 
rejection of the Rindler and other ``noninertial" quantizations 
is in contrast with the general covariance. 

Note first 
that in classical physics the choice of modes for the 
field decomposition has nothing to do with the choice of 
coordinates. In order to describe classical effects as seen by 
a noninertial observer, one can use Minkowski modes as well, but 
expressed in terms of noninertial Fermi coordinates via (\ref{el3}). 
This is what is effectively done in (\ref{corel}) and (\ref{ampl})  
for a quantum case. 

The algebra of quantum fields does not depend on the choice of a 
spacelike hypersurface on which equal-time commutation relations are 
imposed \cite{urb}. However, the knowledge of the algebra does not 
yet determine the physical system; one also needs to specify the 
{\em representation}. The Rindler and the Minkowski quantizations can be 
viewed as two inequivalent representations of the field algebra 
\cite{full,gerlach,fedotov}. Not all possible representations 
need to be realized in nature. Just as Poincare covariance 
related to  
Minkowski spacetime does not imply that tachyons or higher-spin 
particles  
exist, the general covariance does not imply that Rindler 
particles exist.
If the Unruh effect is viewed as in Sec. 5 and  
even if both types of particles exist separately, then,  
contrary to what is often attempted, the 
Rindler quantization cannot serve as an equivalent description
of the Unruh effect. 
All known experiments can be explained in terms 
of Minkowski particles, so it seems that only the Minkowski representation is 
realized in nature.

There is also a controversy in the literature on whether the event 
horizon plays an essential role for understanding the Unruh 
effect. We are now able to give a definite answer to this question. 
If the Unruh effect is treated via the Rindler quantization, then 
the event horizon is essential. If it is treated via a model of a 
particle detector and the Minkowski quantization, then the event horizon is 
not essential. To repeat once more, these two treatments are not 
equivalent.    
       
One could argue that rejecting all representations of 
the field algebra that are not equivalent to the Minkowski 
quantization also excludes quantizations based on 
the replacement of the Fourier integral by a Fourier sum, 
which would be in contradiction with the existence of the 
Casimir effect (see \cite{bd} for a review). However, 
at the fundamental level, the Casimir effect is a nonperturbative 
effect resulting from complicated field {\em interactions}, 
while its description in terms of free fields that vanish 
on the boundary is only an effective, approximative description. 
Standard treatments of the particle production by moving mirrors
(\cite{bd} and references therein) should also be understood in a 
similar way.  

The existence of a preferred representation implies the 
existence of a preferred time, the one with respect to which 
the positive and the negative frequencies, and therefore the lowering and 
the raising operators, are defined. It is important to emphasize 
that this preferred time serves as a tool for the choice of 
the Hilbert space, i.e., the representation of the field algebra, 
but it does not violate the general covariance. It is not 
difficult to accept that in flat spacetime the time of an 
inertial observer has a privileged role (the quantizations based on 
different Lorentz times are equivalent). However, in curved 
spacetime, the quantizations based on Fermi coordinates of different 
inertial observers are not equivalent. How to choose the preferred 
time in general? 

For many theorists, who believe that the fundamental laws of nature 
should be highly symmetrical, it is hard to believe that 
a preferred time could exist. 
Yet, it is an observational fact that a preferred 
time {\em does} exist, the one which is related to the 
homogenity and isotropy of the Universe, as well as to the cosmological 
time arrow, which also seems to coincide with all other time arrows 
\cite{nikolic2}. The time arrow is related to the initial condition 
of the Universe \cite{hawk,nikolic3}. In \cite{nikolic3} it is 
proposed that quantization, i.e., the equal-time commutation relations, 
can also be viewed as an initial condition. The formalism of 
quantum field theory requires a choice of a special time, even 
when functional-integral techniques are employed \cite{nikolic3}. 
In some interpretations of quantum mechanics it is even more explicit 
that a preferred time must exist. For example, if one proposes that 
wave functions collapse instantaneously, one must specify the time 
to which the instantaneousness refers. One cannot propose that this 
refers to the Fermi time of the measuring apparatus, because the 
measuring apparatus can be a large object, so each part of it may 
have a different time. In the de Broglie--Bohm interpretation of 
quantum field theory \cite{holl}, the fundamental, deterministic 
equations of motion possess a preferred time and are manifestly 
noncovariant with respect to general coordinate transformations.   
In the nongeometrical interpretation of gravity,  
which may be the proper approach to quantum gravity, the choice 
of a preferred coordinate frame is also unavoidable \cite{nikolic4}. 
The fixation of a coordinate frame before the quantization 
also resolves the problem of time in quantum gravity \cite{isham}.    
  
Our discussion strongly suggests that there {\em is} a preferred 
coordinate frame, but we still do not know {\em which} coordinate  
frame is the right one. Ultimately, 
this question should be answered 
experimentally. However, it is natural to expect that there is 
only one preferred time, i.e., that the time related to the quantization 
is the same time that is relevant to cosmology and the time arrow. 
This naturally leads to the conclusion that the right coordinates  
could be the normal Gauss coordinates, for which the metric takes the form 
\begin{equation}\label{gauss}
ds^2 =dt^2-g_{ij}dx^i dx^j \; .
\end{equation}
These coordinates are a generalization of Minkowski coordinates 
to curved spacetime.  
These also correspond to the simplest possible choice of coordinates
in the Arnowitt-Deser-Misner approach to canonical gravity \cite{adm}. 
It seems that these coordinates could allow a global quantization 
for the whole Universe. For example, if the topology of the Universe is 
${\bf R}\times S^3$, where ${\bf R}$ represents the time, 
then one needs at least two regions with different 
coordinates. However, these two coordinate frames may be related by a 
transformation of the form $t'=t$, $x'^i =f^i (x^1,x^2,x^3)$, which 
does not influence the time coordinate, which is essential for 
quantization. The Gauss coordinates are also appropriate for a global 
quantization in a black-hole background, because the metric is 
given by \cite{mol}
\begin{equation}\label{mol1}
ds^2= dt^2 -\frac{2M}{r_s}dr^2 -r_s^2 d\Omega^2 \; ,
\end{equation}
where
\begin{equation}\label{mol2}
r_s =(9M/2)^{1/3}(r-t)^{2/3} 
\end{equation}
is the Schwarzschild radial coordinate, 
so (\ref{mol1}) does not possess any coordinate singularity, but 
only the genuine 
singularity at $r_s =0$. 

We do {\em not} require that the representations at different 
values of the {\em prefered} time should be equivalent. Therefore,  
when $g_{ij}$ in (\ref{mol1}) depends on $t$, the particle 
creation is possible. This allows the particle creation by 
the Universe expansion, but does not allow the Unruh effect 
via the Rindler quantization.   
It would be interesting to study 
Hawking radiation starting from the quantization with respect to 
the time $t$ in (\ref{mol1}), but the standard prediction of the 
thermal spectrum \cite{hawk2} is unlikely to be altered, because, 
if the radiation is viewed merely as particles
that escaped from a collapsing body just before the event horizon
was formed \cite{bd}, the 
thermal spectrum can be understood even with classical physics 
\cite{padm4,padm5}.   

The preferred time leads to the preferred representation of the 
field algebra and therefore to the unambiguous formal definition 
of particles in terms of raising and lowering operators. 
There is also a simple,     
operational argument, why the number of real particles in a given 
state does not depend on the motion of the observer. Quantum detectors 
may respond to real particles as well as to ``virtual particles" 
(or, better to say, to quantum fluctuations). On the other hand, 
classical detectors respond only to real particles. We believe 
that it is quite clear that two classical detectors will detect 
the same number of particles, irrespective of their motion  
(of course, providing that the trajectory of each particle crosses 
the trajectory of each detector once and only once). If we define  
radiation as emission of real particles, in this way we also confirm the 
conclusion of Sec. 4 that radiation does not depend on the observer.  
  
\section{Conclusion}

For a given coordinate frame, values of covariant
fields are mathematically determined at 
all spacetime points. However, a local observer can measure only
the values at the point of his own position.    
By exploring the coordinate transformation between two
Fermi frames, we have shown that two observers, having the same instantaneous
position and velocity, will observe the same values of covariant fields
at their common instantaneous position, even if they have different
instantaneous accelerations. In particular,
this implies that in classical physics
the notion of radiation is observer independent. 
Consequently, an inertial charge in flat
spacetime does not radiate from the point of view of an accelerated
observer. However, a ``freely" falling charge in
{\em curved} spacetime does not move along a geodesic, so it
{\em does} radiate (as seen by any observer).

The Rindler-quantization approach and the 
``particle-detector" approach to the Unruh effect are not 
equivalent. The Rindler quantization corresponds to a 
representation of the field algebra which is not equivalent 
to the Minkowski quantization, so the 
Rindler quantization should be rejected. 
There is a strong evidence that a preferred coordinate frame 
must exist, which allows an unambiguous formal definition 
of the particle number in curved spacetime. The response 
of classical particle detectors leads to an operational 
argument that the particle number does not depend on 
the observer.

Our discussion suggests that the fundamental problems of 
cosmology and quantum field theory are closely related. 
However, the concept of particle in quantum field theory is 
still far from being completely clear, 
even in Minkowski spacetime. For a clearer picture,  
we believe that it is essential to 
have a better understanding of nonperturbative 
effects, density-matrix decoherence, and the wave-function collapse.

\section*{Acknowledgment}   

This work was supported by the Ministry of Science and Technology of the
Republic of Croatia under Contract No. 00980102.

\end{document}